\documentclass[12pt]{iopart}

\usepackage{graphicx}
\begin{document}

\title[STAR results on medium properties .....]{STAR results on medium properties and response of the medium to energetic partons}

\author{Bedangadas Mohanty(for the STAR Collaboration)}

\address{Variable Energy Cyclotron Centre ,Kolkata-700064 ,INDIA.}
\ead{bmohanty@veccal.ernet.in}
\begin{abstract}
We report new STAR results on the consequences of highly energetic partons 
propagating through the medium formed in heavy ion collisions using correlations 
as an experimental probe. The recent results providing insights about color factor effects 
and path length dependence of parton energy loss, system size dependence of di-hadron 
fragmentation functions, conical emission and ridge formation in heavy ion collisions are presented.

\end{abstract}


\section{Introduction}
After the discovery of the phenomenon of jet quenching at RHIC, the focus has been to study 
the properties of the medium formed in heavy ion collisions at high energy~\cite{whitepaper}.  The basic approach 
has been to use the highly energetic partons available in the initial stages of the collisions as a 
probe for the properties of the subsequent medium that is formed. These partons then fragment into jets 
of hadrons and we look for modification to the properties of these jets. This is mostly done at RHIC by 
comparing the various experimental observations in $p$+$p$, $d$+Au and Au+Au collisions. The properties of the 
medium which we would like to study include, the energy density achieved in the collisions and the velocity of 
sound in the medium. In addition, we would like to understand better the role of interactions between the 
constituents and the degree of collectivity and thermalisation of the matter. 
It is found that correlations between the produced particles in the heavy ion collisions can be 
used as an experimental tool to address these issues. The study of the medium through high $p_{T}$ triggered 
correlations is the main theme of this article. Only a few selected topics discussing results on parton energy loss
in the medium and the response of the medium to passage of energetic partons are covered in this article. 
More details can be obtained from STAR presentations at Quark Matter 2008~\cite{starqm}.

\section{Parton energy loss in the medium formed in heavy ion collisions}

The non-Abelian nature of quantum chromodynamics (QCD) results in the gluons losing more energy than quarks 
in the medium formed in high energy heavy-ion collisions. Experimental results in $p$+$p$ collisions when compared 
to NLO pQCD calculations show that at high transverse momentum ($p_{T}$) the produced $p+\bar{p}$ are 
dominantly from gluon jets and charged pions have substantial contribution from quark jets. If such a scenario 
is applied to heavy-ion collisions at RHIC, one would expect the difference in quark and gluon energy loss to 
have an effect on measured observables, such as   $\bar{p}/p$ ratio at high $p_{T}$ and the nuclear modification 
factor ($R_{AA}$) for $\pi^{+}+\pi^{-}$ and $p+\bar{p}$.

\begin{figure*}[htp]
\begin{center}
\includegraphics[scale=0.38]{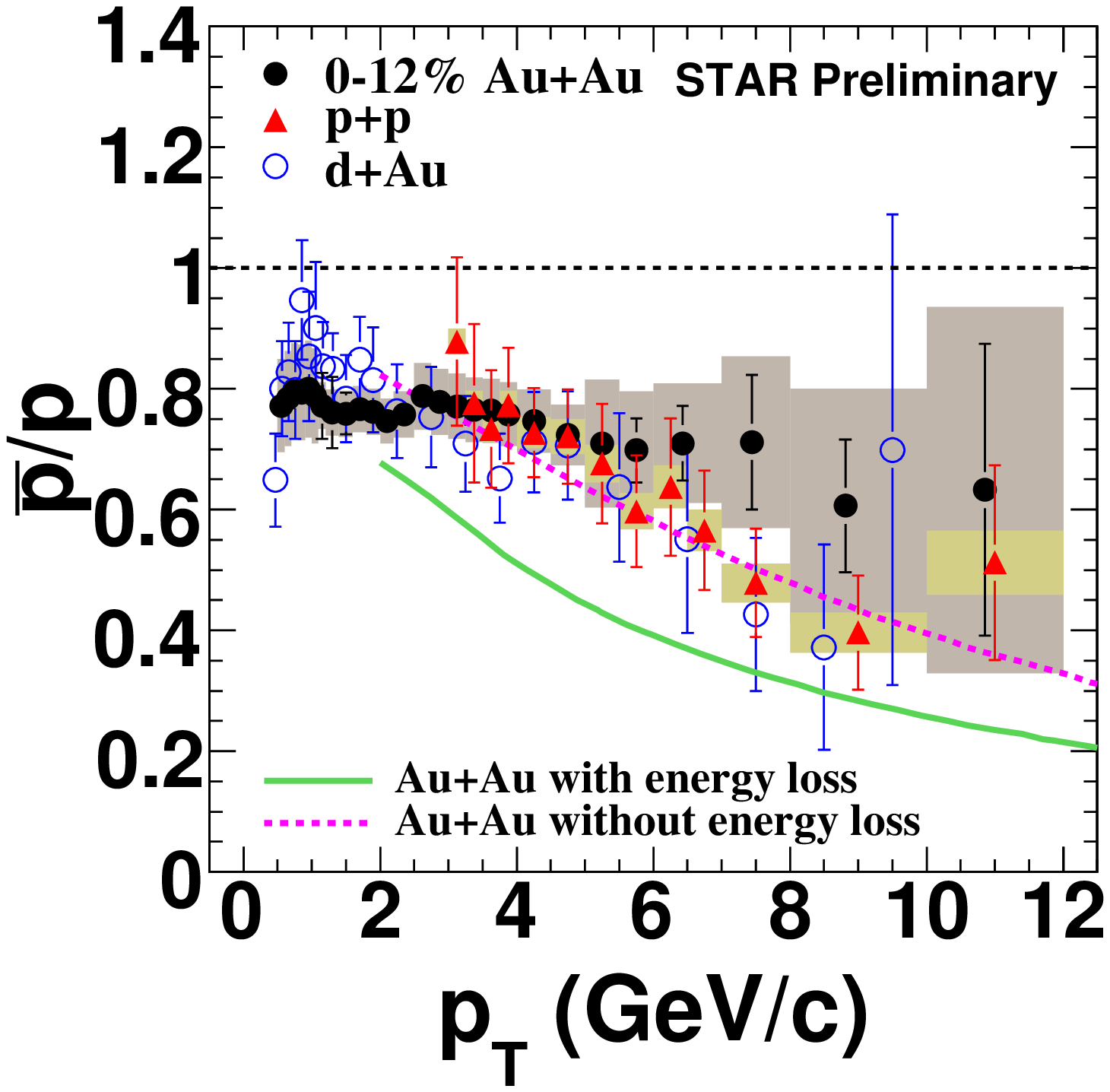}
\includegraphics[scale=0.38]{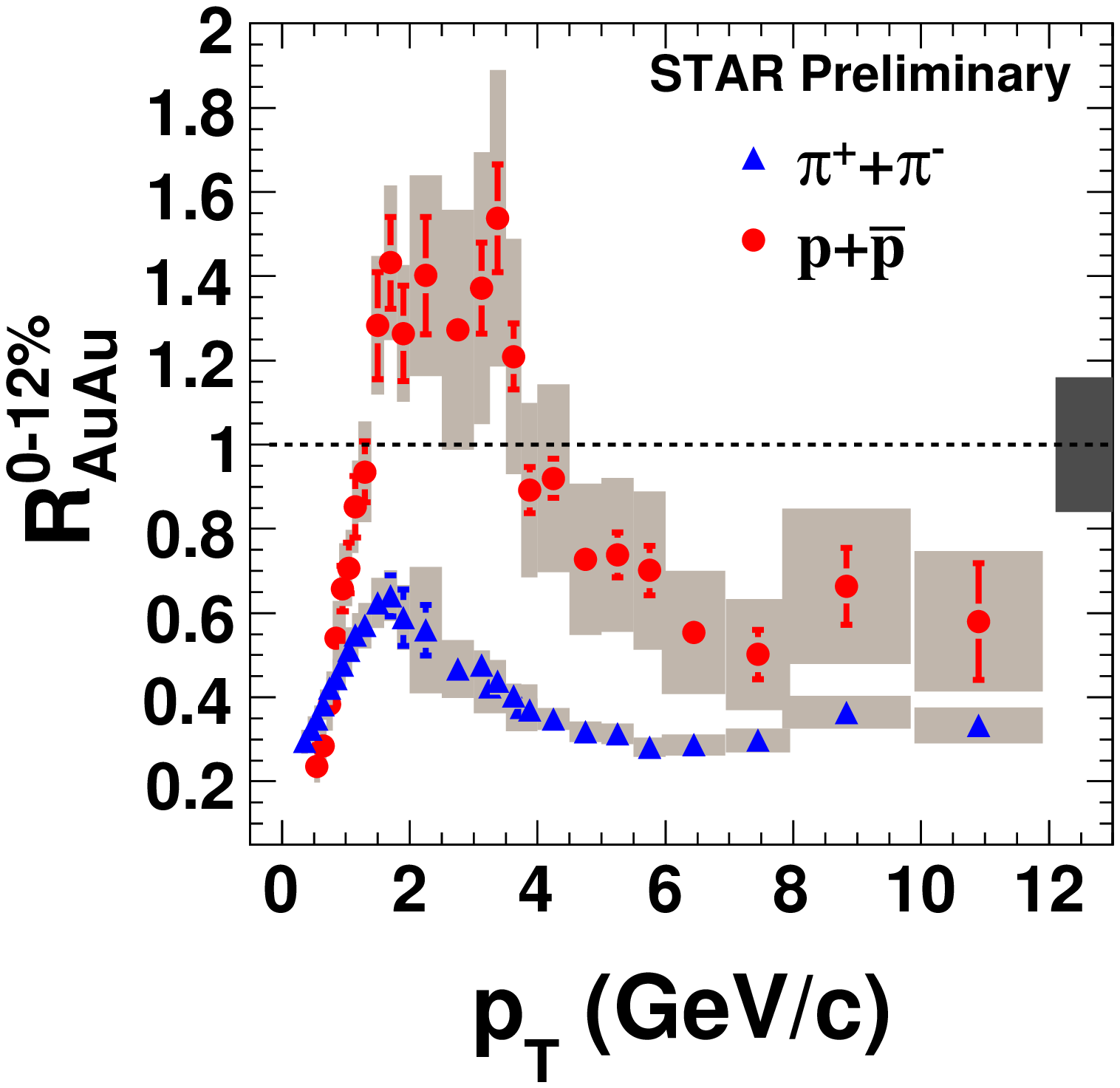}
\caption{Left panel : $\bar{p}/p$ ration in central Au+Au, minimum bias $d$+Au and $p$+$p$ collisions at 200 GeV. 
The data are compared to model calculations without energy loss (dashed line) and with the color factor 
effects in parton energy loss (solid line)~\cite{thcolor}. Right panel: Transverse momentum dependence of 
the nuclear modification factor  for $\pi^+ + \pi^-$ and $p+\bar{p}$ in central Au+Au collisions.}
\label{color}
\end{center}
\end{figure*}

Fig.~\ref{color}(left) shows that the $\bar{p}/p$ ratio for central Au+Au collisions at 200 GeV  
at high $p_{T}$ ($>$ 6 GeV/$c$) is comparable or slightly higher than the $p$+$p$ and 
$d$+Au results~\cite{starcolor}. This is in contrast to the naive expectations from color charge dependence of 
parton energy loss, where the ratio at high $p_{T}$ in Au+Au collisions is expected to be lower than $p$+$p$ collisions. 
Comparison to model calculations without energy loss are in reasonable agreement with the $p$+$p$ and $d$+Au results, 
whereas calculations including color charge dependence of energy loss~\cite{thcolor} give a much lower value of the 
$\bar{p}/p$ ratio compared to data for most of the measured $p_{T}$ range. The stronger coupling of 
the gluons with the medium formed in Au+Au collisions is expected to lead to a lower value of the $R_{AA}$ 
an experimental observable that reflects parton energy loss.  
Fig.~\ref{color}(right) shows the $R_{AA}$ for $p$+$\bar{p}$ (expected to be dominantly from gluons )  
is comparable or slightly higher than for  $\pi^{+}+\pi^{-}$ (expected to have substantial contribution from quarks) 
at high $p_{T}$ ($>$ 6 GeV/$c$) for central Au+Au collisions at  200 GeV~\cite{patricia}. This is in contrast to the naive 
expectation of a lower $R_{AA}$ for $p$+$\bar{p}$ compared to $\pi^{+}+\pi^{-}$.

\begin{figure*}[htp]
\includegraphics[scale=0.26]{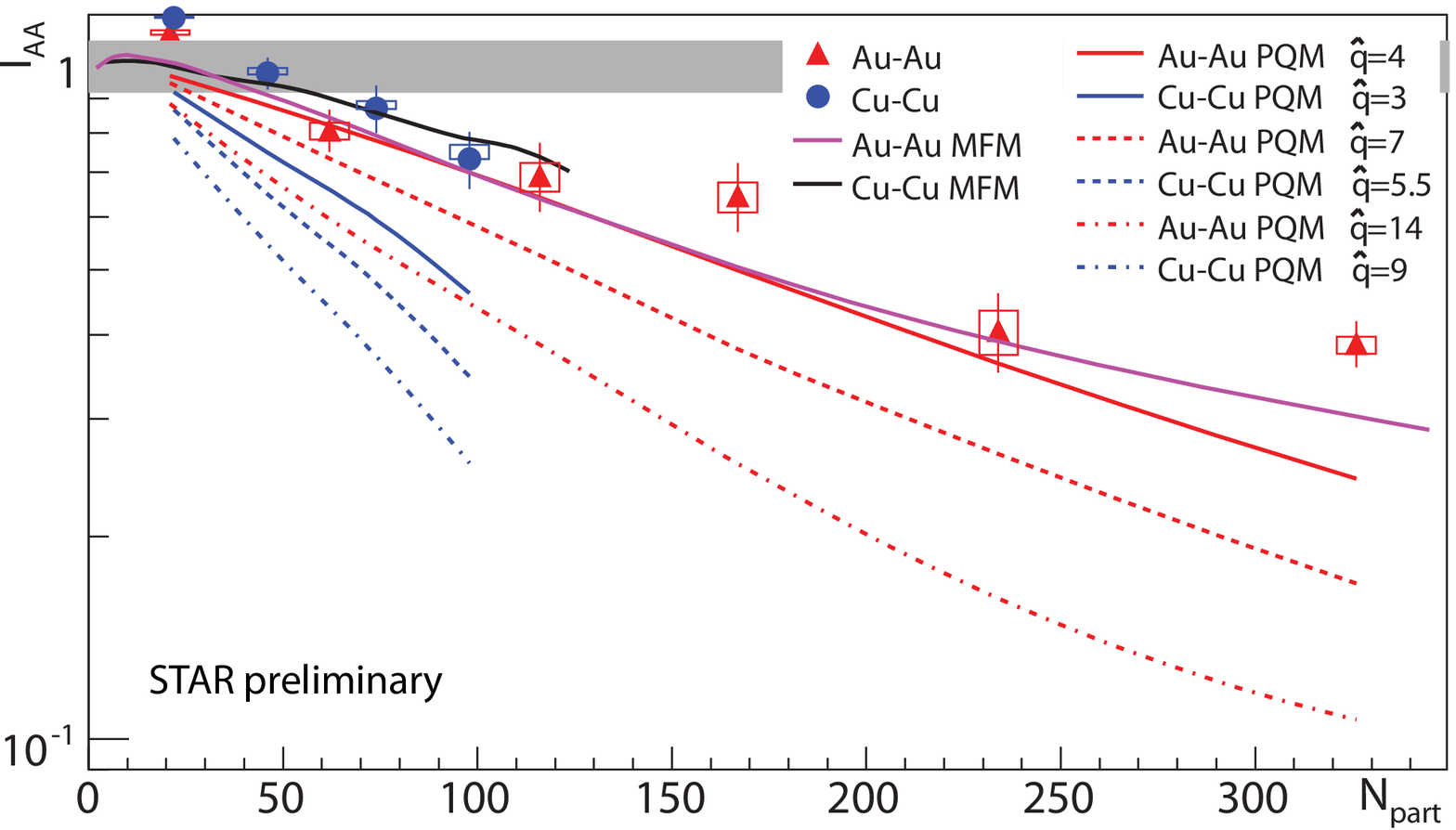}
\includegraphics[scale=0.26]{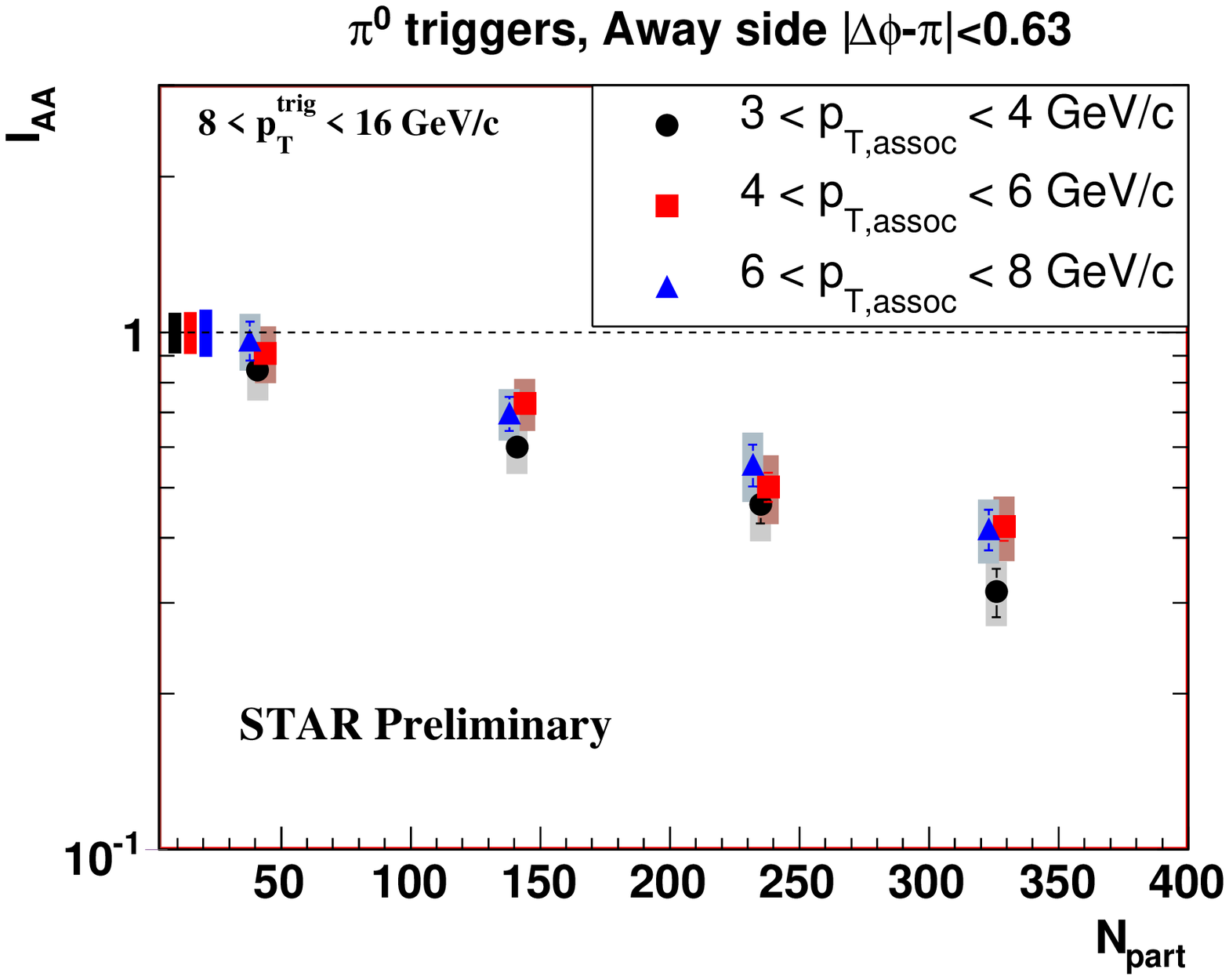}
\includegraphics[scale=0.26]{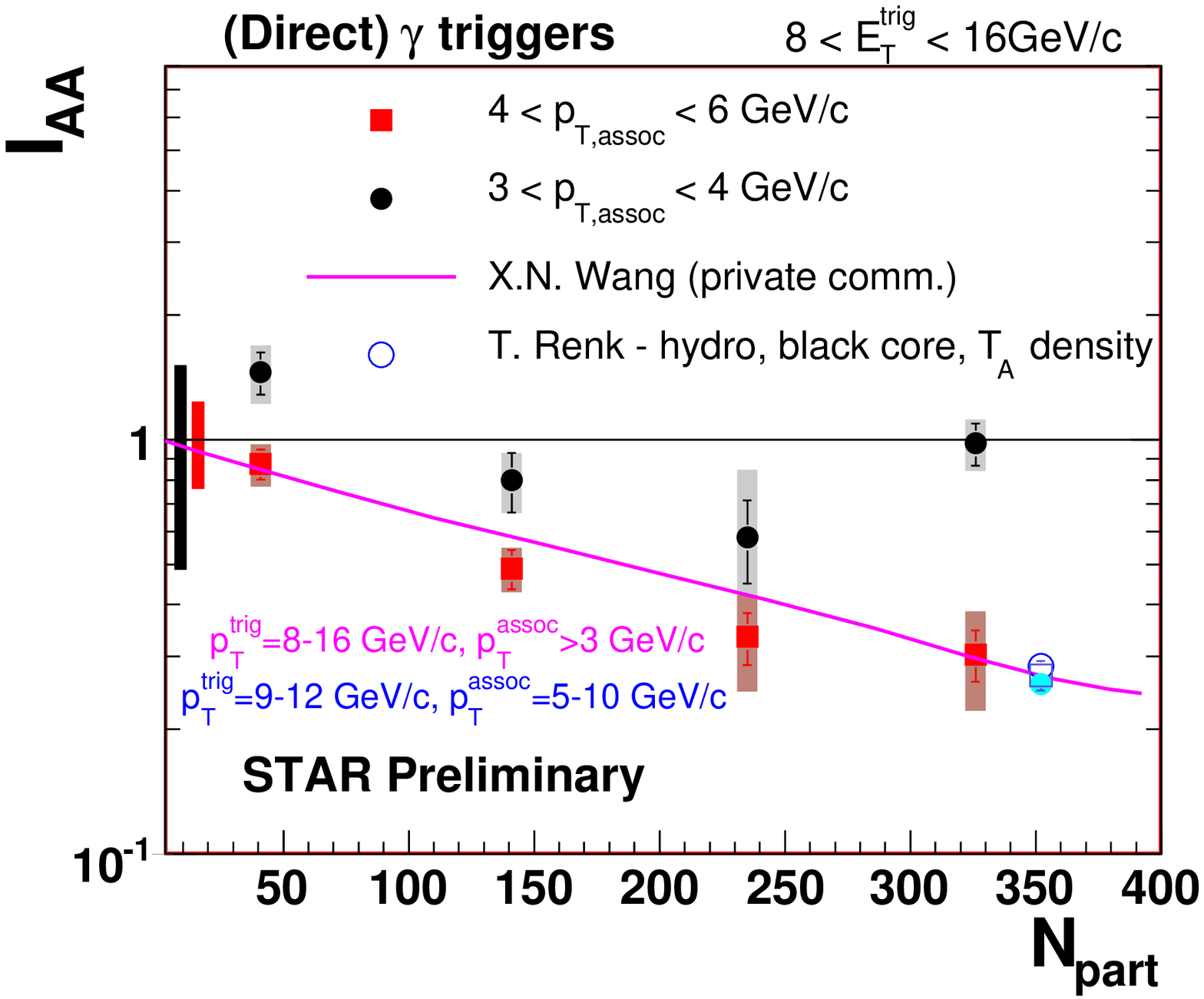}
\caption{Left panel : Centrality dependence of  di-hadron  $I_{AA}$ in Cu+Cu and Au+Au  relative to d+Au collisions 
at 200 GeV. The results are compared to PQM~\cite{pqm} (blue and red lines) and MFM~\cite{mfm} (magenta and black lines) calculations.
Middle panel : Centrality dependence of $\pi^{0}$-trigger $I_{AA}$ in Au+Au relative to p+p collisions at 
200 GeV for various $p_{T,assoc}$ ranges. Right panel : same as middle panel but for $\gamma$-triggers 
and compared to theoretical calculations~\cite{renk,wang}.}
\label{eloss}
\end{figure*}

Although the nuclear modification factor for produced particles reflects the energy loss of partons in 
the medium formed in heavy-ion collisions, were found to have a limited sensitivity to different 
mechanisms of partonic energy loss in the medium~\cite{renk}.  It was suggested that studies of di-hadron 
correlations can provide better sensitivity to properties of the medium~\cite{mfm}. In STAR we have carried out
a systematic study of di-hadron correlations with various collision species at different beam energies 
as a function of collision centrality~\cite{oana}. One such result of $I_{AA}$ is shown in the Fig.~\ref{eloss}(left panel) 
for 200 GeV Au+Au and Cu+Cu collisions as a function of number of 
participating  nucleons ($N_{part}$). The $I_{AA}$ is  defined as the integrated yield of the away-side 
associated particles per high $p_{T}$ trigger particle scaled by the corresponding yield measured in d+Au collisions.  
The $N_{part}$ scaling which is seen in the data is reproduced by the Modified 
Fragmentation Model (MFM)~\cite{mfm},
while the Parton Quenching Model (PQM)~\cite{pqm} shows no such scaling. Further comparison of these models 
with data should  provide better insight in the energy loss mechanism and the path-length dependence of energy loss.

It was further suggested that studying $\gamma$-hadron correlations would be more sensitive to the different 
mechanisms of partonic energy loss and also provide full accounting of the jet energy loss~\cite{renk}.
Fig.~\ref{eloss} (middle panel) shows the $I_{AA}$ with $\pi^{0}$ triggers. 
The $I_{AA}$ values $<$ 1 for central collisions  supports the earlier picture of jet quenching at RHIC. 
Fig.~\ref{eloss}(right panel) shows the first measurements of $I_{AA}$ from $\gamma$-hadron correlations 
in STAR as a function of $N_{part}$ in Au+Au collisions at 200 GeV~\cite{ahmed}. The $I_{AA}$ shown in the 
figure is defined as the integrated yield of the away-side associated particles per direct photon trigger 
scaled by the corresponding yield measured in $p$+$p$ collisions.
The results for $\gamma$-triggers 
are with 8 $<$ $E_{T}$ $<$ 16 GeV/$c$ and the associate hadrons within 3 $<$ $p_{T,assoc}$ $<$ 4 GeV/$c$  
and 4 $<$ $p_{T,assoc}$ $<$ 6 GeV/$c$ ranges. 
Within the current uncertainty the $I_{AA}$ agrees with the theoretical calculations~\cite{renk,wang}. With future luminosities at 
RHIC II, we will be sensitive at the level to distinguish theoretical mechanisms.
 
\section{Response of the medium to the passage of energetic partons}

It has been previously reported at RHIC, that on the away side of a high $p_{T}$ trigger particle, the 
correlated yields are strongly suppressed at $p_{T}$ $>$ 2 GeV/$c$, while at lower $p_{T}$ 
the yield is enhanced~\cite{dihadron}. Further in the away side the correlated hadrons appear to be partially 
equilibrated with the bulk medium and the distributions are broader (sometimes double peaked) in azimuth~\cite{dihadron2}. 
Identifying the underlying physics mechanism for such observations, one of which could be due to Mach cone 
shock-waves being generated by a large energy deposition in a hydrodynamic medium~\cite{thmach}, can provide important 
information about properties of the medium, such as the speed of sound and equation of state. 
3-particle azimuthal correlations between a high $p_{T}$ trigger particle and two associated particles 
have the ability to discriminate the different underlying physics mechanisms.

\begin{figure*}[htp]
\includegraphics[scale=0.38]{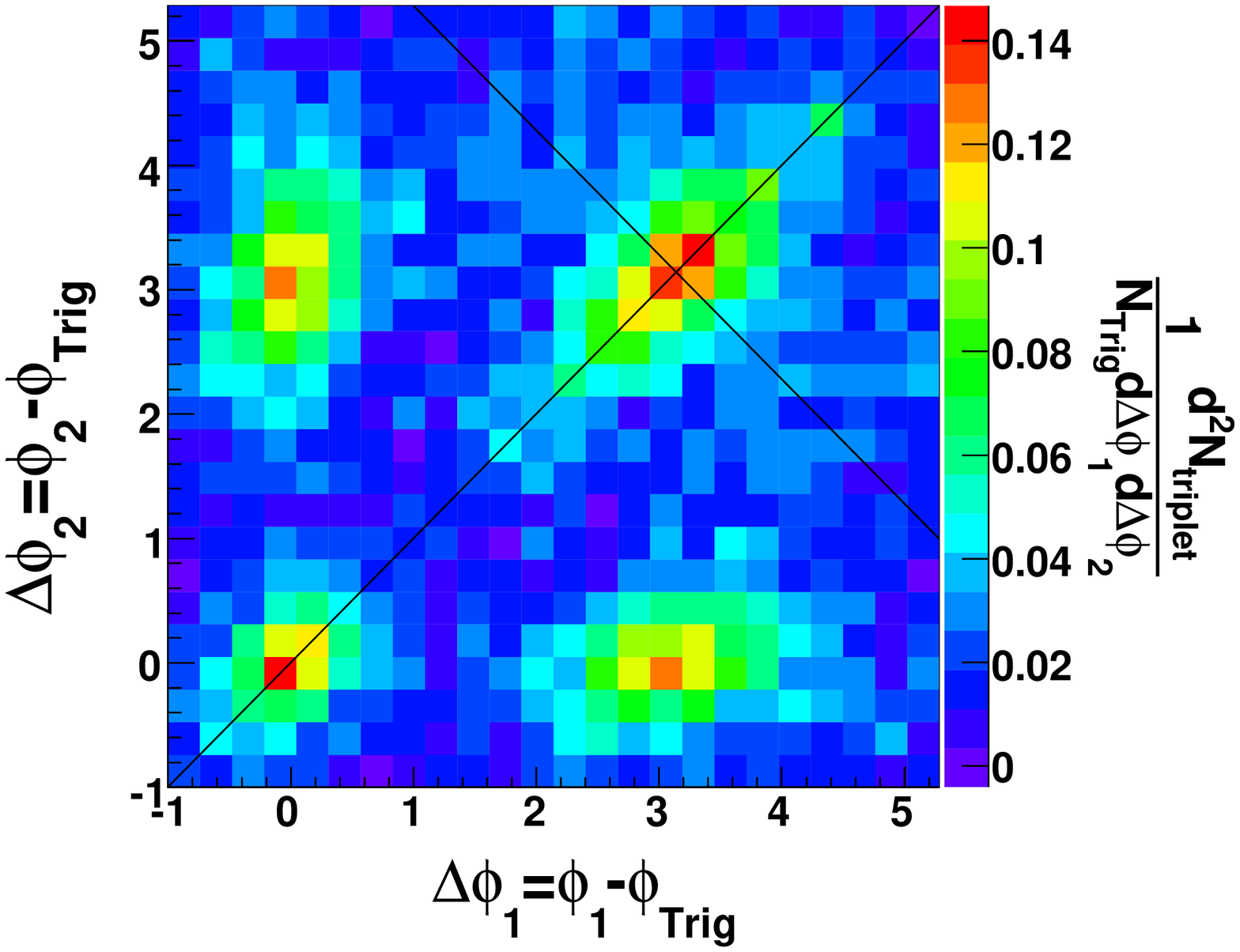}
\includegraphics[scale=0.38]{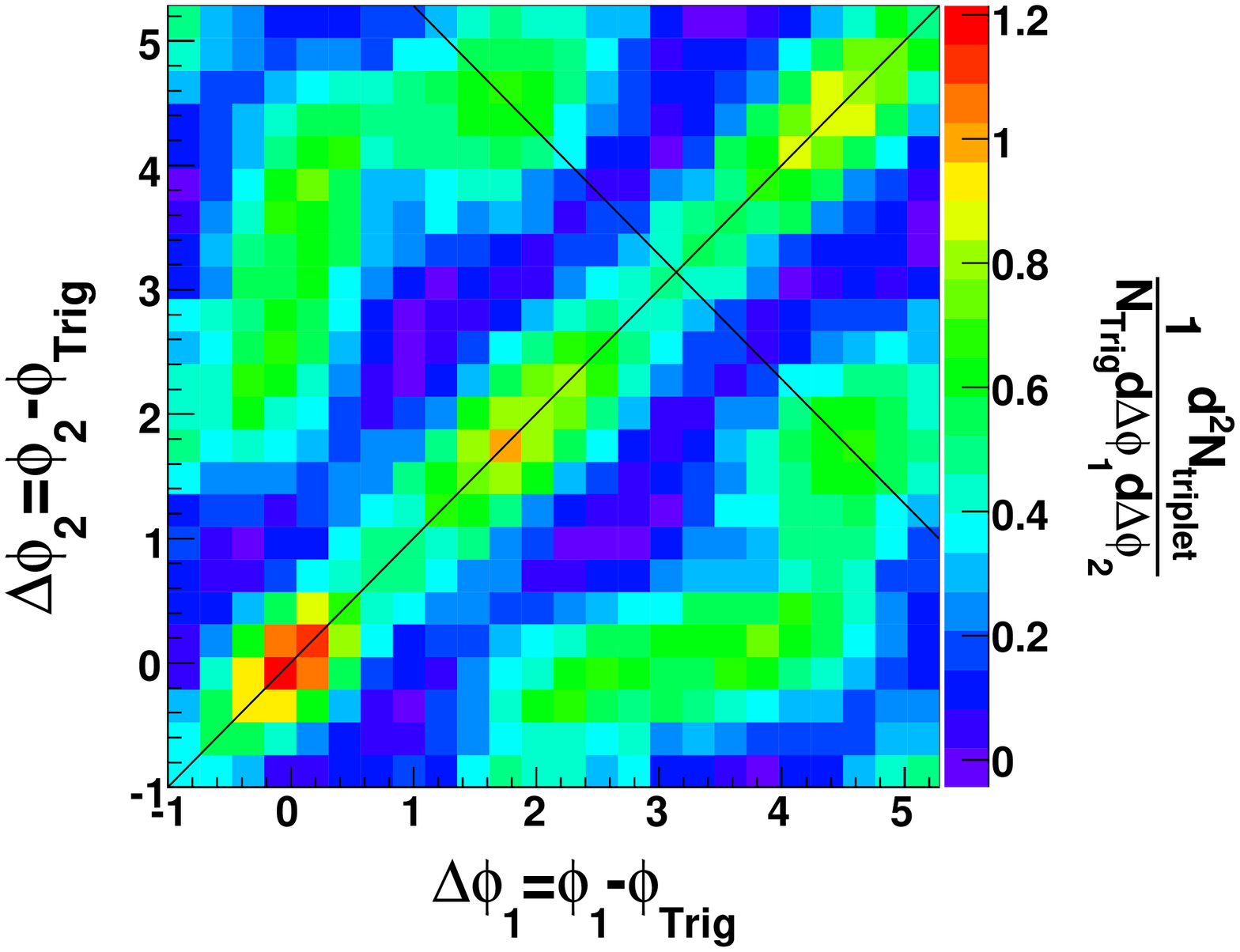}
\caption{Preliminary 3-particle azimuthal angle correlations for minimum bias d+Au collisions (left panel) and 0-12\% 
central Au+Au collisions (right panel) at 200 GeV. The trigger particles have $p_T$~=~3-4~GeV/$c$ and 
associated particles have $p_T$~=~1-2~GeV/$c$.}
\label{conical}
\end{figure*}

Fig.~\ref{conical} shows the background subtracted 3-particle azimuthal angle correlations in minimum bias d+Au (left) 
and 0-12\% central Au+Au (right) collisions at 200 GeV. The trigger particles are in the range 
3 - 4 GeV/$c$ in $p_{T}$ and associated particles in the range 1- 2 GeV/$c$ in $p_{T}$. The analysis assumes 
the event to be composed of two components, one that is correlated with the trigger and the other is 
background, uncorrelated with the trigger except the indirect correlations via anisotropic flow~\cite{jason}. 
In Au+Au collisions we observe that opposite to the trigger particle in azimuth, the associated particles 
seem to populate preferentially on a cone around $\pi$ radians, being equally far apart and symmetric about $\pi$ 
and close together.  Distinct peaks around angles of $\sim$ 1.4 radians (cone angle) away from $\pi$ 
are only observed 
in Au+Au collisions and not in d+Au. It may be noted that in d+Au system neither a large partonic energy deposition 
is observed nor the medium is expected to show substantial collectivity. These additional structures in Au+Au 
collisions are clear experimental evidence of conical emission of correlated hadrons with high $p_{T}$ particles. 
Further we observe that the cone angle is independent of collision centrality and the associated particle $p_{T}$, 
lending a support to the picture of Mach cone shock waves propagating in central heavy-ion collisions~\cite{jason}. 
A separate analysis in STAR based on 3-particle cumulants~\cite{claude} however has so far not seen a clear evidence of 
such a conical emission.

\begin{figure*}[htp]
\begin{center}
\includegraphics[scale=0.25]{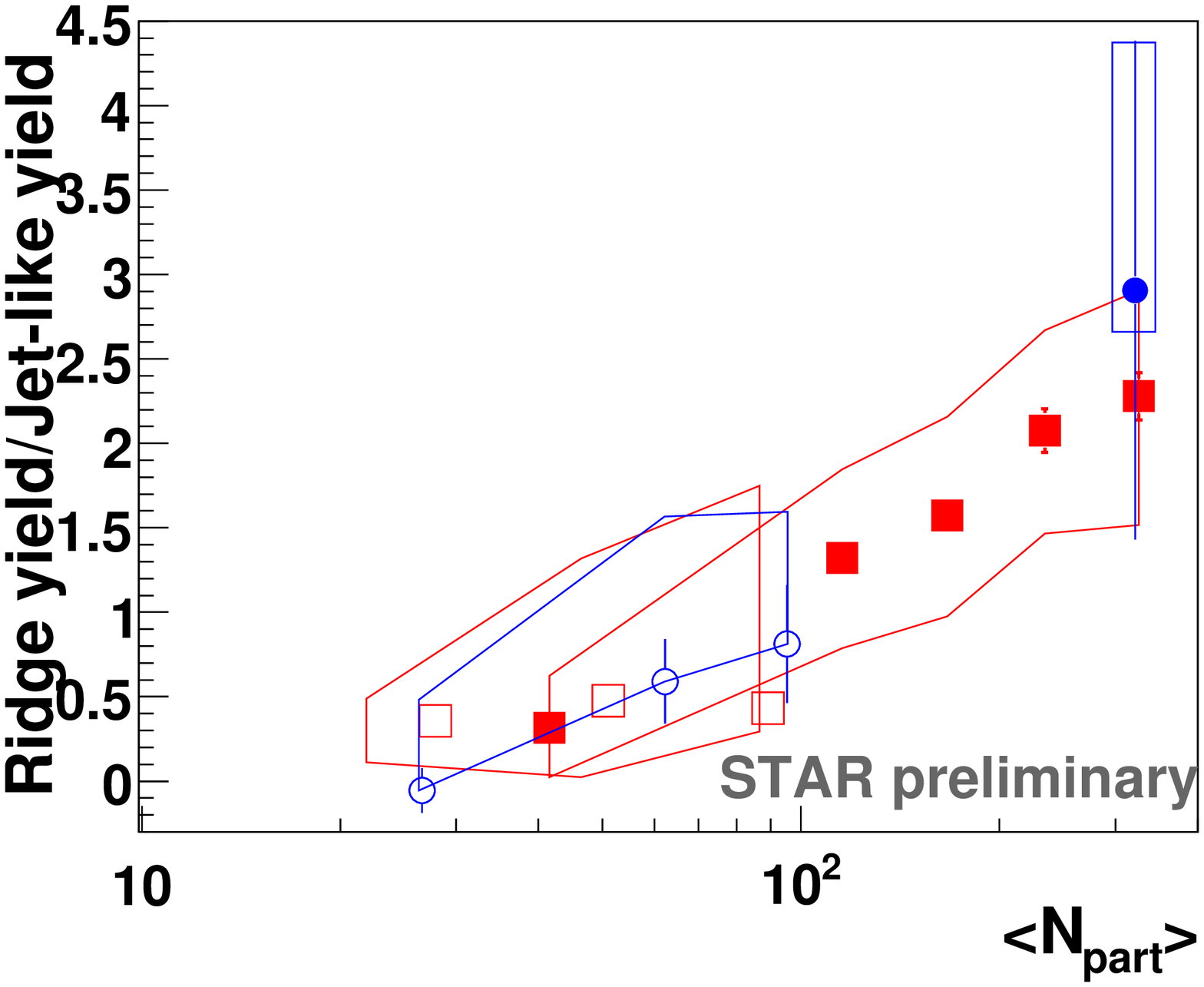}
\includegraphics[scale=0.25]{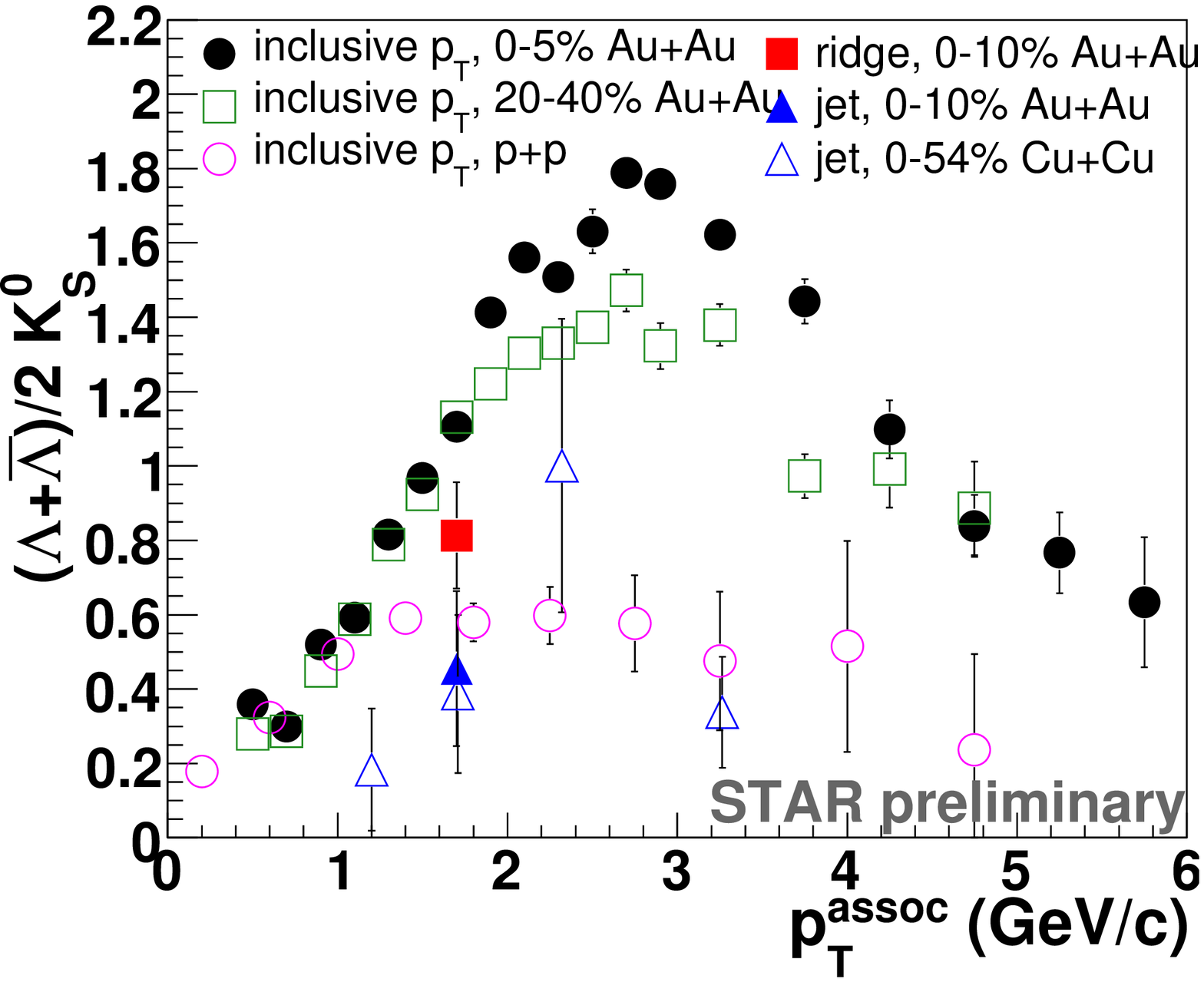}
\includegraphics[scale=0.25]{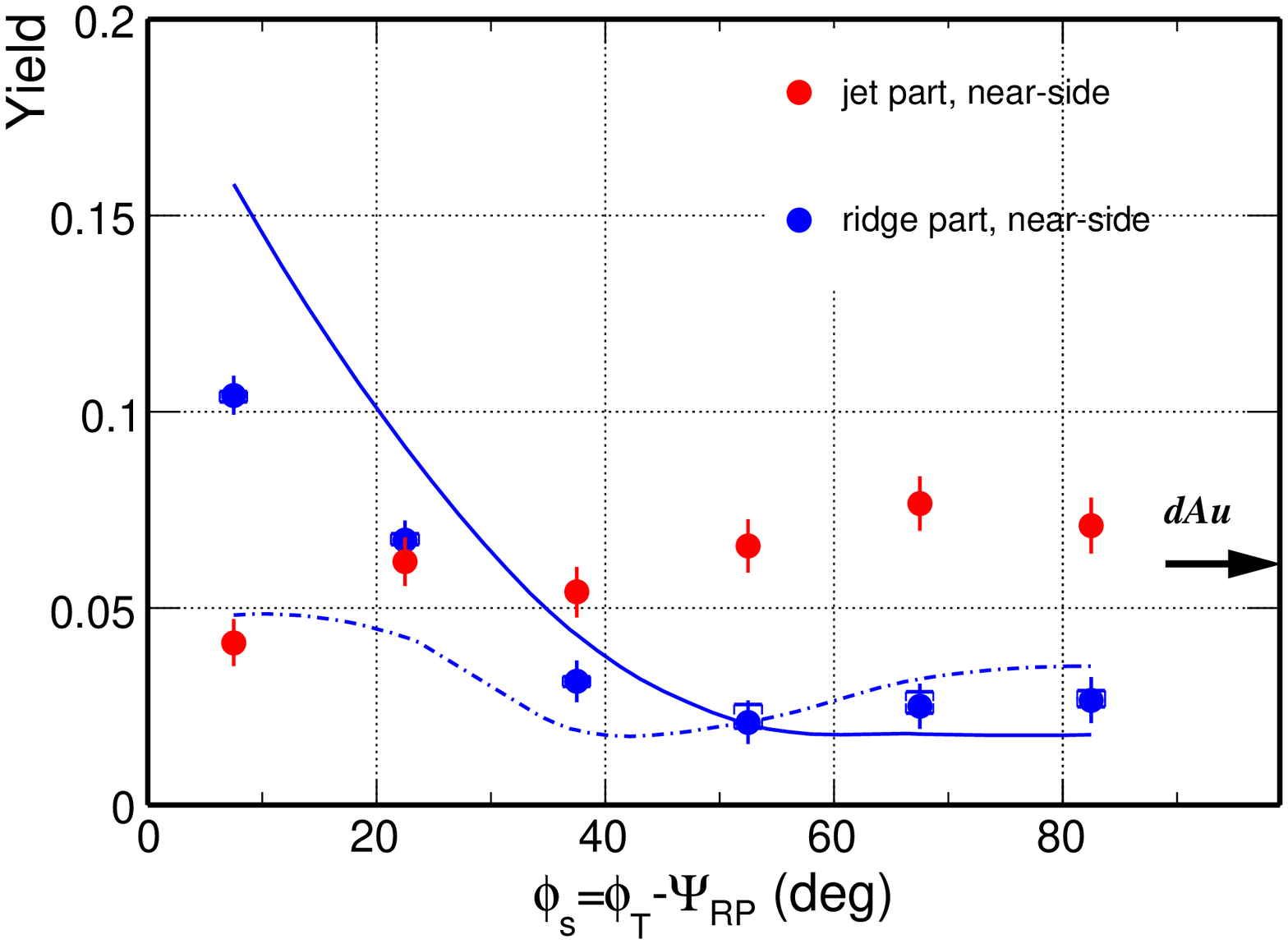}
\caption{Left panel : $N_{part}$ dependence of ridge/jet yield ratio in Au+Au (solid symbols) 
and Cu+Cu (open symbols) collisions at 62.4 GeV(circles) and 200 GeV (squares).
Middle panel : $(\Lambda + \bar{\Lambda})/2K^{0}_{S}$ as a function 
of $p_{T}$ for inclusive, ridge and jet regions in Au+Au and Cu+Cu collisions at 200 GeV.
Right panel : Ridge and jet yields from di-hadrons correlations in 
non-central Au+Au collisions at 200 GeV as  a function of angular difference between 
the trigger particle and the reaction plane. The lines are systematic uncertainties due to elliptic flow subtraction.}
\label{ridge}
\end{center}
\end{figure*}

\begin{table}
\caption{\label{table1}
Observations in Ridge and Jet cone regions of high $p_{T}$ triggered di-hadron correlations in near side for 
heavy ion collisions }
\begin{center}
\begin{tabular}{|c|c|c|}
\hline
Observable & Observation in ridge & Observation in jet cone  \\
\hline
Inverse slope of $p_{T}$ spectra ($T$)   & $T_{ridge}$ $\sim$ $T_{inclusive}$      &  $T_{jet}$ $>$ $T_{inclusive}$ \\
\hline
$N_{part}$ dependence  & Increases with $N_{part}$               & Constant with $N_{part}$ \\
  of yields ($Y$)                                     &                                         & \\
\hline
$\sqrt{s_{NN}}$ dependence         & Y(62.4) $<$ Y(200)  & Y(62.4) $<$ Y(200) \\
 of yields                            &                                         & {\bf Ridge/Jet ratio} similar  \\
                            &                                         & at both energies (Fig.~\ref{ridge})\\
\hline
Collision system  & For a $\sqrt{s_{NN}}$        &   For a $\sqrt{s_{NN}}$ \\
dependence of yields    &  \& similar $N_{part}$,                  &   \& similar $N_{part}$, \\
    &                      Y(Au+Au) $\sim$ Y(Cu+Cu)                      & Y(Au+Au) $\sim$ Y(Cu+Cu)               \\
\hline
Trigger particle type  & No significant           & No significant  \\
dependence of    &       dependence                              & dependence  \\
yields per trigger particle                                 &                                    & \\
\hline
Baryon/meson ratios  (Fig.~\ref{ridge})      & Similar to inclusive               & Smaller than inclusive \\
\hline
\end{tabular}
\end{center}
\end{table}
Studies of near-side di-hadron correlations ($\Delta\phi$ $\sim$ 0) at high $p_{T}$ $>$ 6 GeV/$c$ revealed a 
jet-like correlation at small pair phase space separation ($\Delta\phi$ $\sim$ 0, $\Delta\eta$ $\sim$ 0) which is 
unmodified in central Au+Au collisions relative to d+Au, suggesting that the dominant production mechanism is 
jet fragmentation outside the dense medium. At lower momentum, significant correlated yield has been observed 
in central collisions at large pair separation in pseudo-rapidity $\Delta\eta$ (the {\bf ridge})~\cite{dihadron2,joern}. A feature
observed in Au+Au collisions and not observed in $d$+Au collisions.
However, inclusive hadron production at moderate $p_{T}$ $<$ 6 GeV/$c$ in central Au+Au collisions 
differs significantly from $p$+$p$ and $d$+Au collision systems, and jet fragmentation may not be the 
dominant hadron production mechanism in this region~\cite{starcolor}. It is therefore an open question whether 
the ridge is a novel manifestation of partonic energy loss~\cite{armesto} or is due to a different, 
non-perturbative mechanism such as recombination~\cite{hwa}. In order to provide further experimental 
inputs for deciphering the mechanism for ridge formation, STAR reported at the QM2008 several new 
results~\cite{nattrass,aoqi}. Most of the results presented make the assumption, based on experimental 
observation, that the near-side projection onto $\Delta\eta$ can be separated into a jet-like 
peak centered at  $\Delta\phi$ $\sim$ 0, $\Delta\eta$ $\sim$ 0 and an independent ridge component. 
Then we have studied the properties of various observables in these two components which are given in Table~\ref{table1}
and Fig.~\ref{ridge}.

In addition to the above experimental observations in the near side from high $p_{T}$ triggered 
di-hadron correlations, STAR has carried out a detailed study of the near side triggered 
di-hadron correlations with respect to 
reaction plane~\cite{aoqi}. In Au+Au collisions at 200 GeV, it is observed that the ridge yield seems to decrease 
with increase in the azimuthal angle difference between the trigger particle and the reaction plane 
whereas the jet yield is constant as the function of the angular difference and similar to that observed 
in d+Au collisions. This observation (right panel of Fig.~\ref{ridge}) can be interpreted as strong near-side 
jet-medium 
interaction in reaction plane possibly generating the sizable ridge and there is minimal near-side 
jet-medium interactions perpendicular to the reaction plane.

\begin{figure*}[htp]
\begin{center}
\includegraphics[scale=0.78]{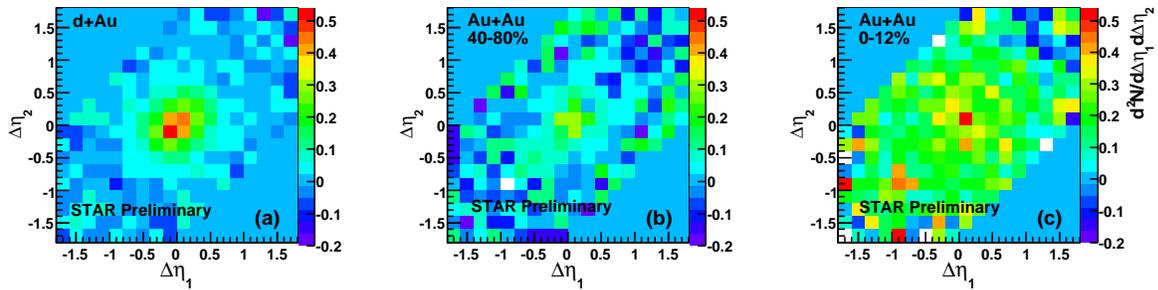}
\caption{3-particle correlations in pseudo-rapidity for small azimuthal angle difference between associated particles and trigger particle for minimum bias d+Au collisions and Au+Au collisions  at 200 GeV.  
The central red box indicates the jet-peak structure, observed in both d+Au and Au+Au collisions.}
\label{etaridge}
\end{center}
\end{figure*} 
STAR has also carried out 3-particle correlations in $\Delta\eta-\Delta\eta$~\cite{pawan} 
and the results are shown in Fig.~\ref{etaridge} 
for minimum bias d+Au collisions (left) and central Au+Au collisions (right) at 200 GeV. The trigger particle 
$p_{T}$ range being 3--10 GeV/$c$ and associated particle $p_{T}$ range is between 1--3 GeV/$c$. A clear jet peak 
is observed in these correlations at $\Delta\eta-\Delta\eta$ $\sim$ 0 for both d+Au and Au+Au collisions. 
In addition, an uniform over all excess of associated particles is observed in Au+Au collisions. 
This observation is contrary to most model predictions, which would lead to an excess of yield along 
the diagonals or strips across the $\Delta\eta_{1}$ = 0 and $\Delta\eta_{2}$ = 0.

\section{Conclusions}
In conclusion, the theoretically expected differences in energy loss between quarks and gluons 
are not experimentally observed in the measured $p_{T}$ range for Au+Au collisions at RHIC.	
First measurements of $\gamma$-hadron correlations in Au+Au collisions at 200 GeV are reported.
In future, detailed comparison with model calculations should be able to provide insights into 
the mechanism of parton energy loss in the medium formed in heavy ion collisions.	
Strong jet-medium interactions have been observed in Au+Au collisions at RHIC.  
Signals of conical emission are observed in central Au+Au collisions at 200 GeV from a 
detailed analysis based on a 2-component approach. The Medium seems to respond through formation 
of ridge. New observations such as particle ratios in the ridge are similar to inclusive, 
ridge being dominated in-plane and 3-particle correlations in $\Delta\eta$ reflecting a physics 
scenario of jet fragmentation along with an uniform overall excess of associated particles, 
should now provide more stringent constraints on the proposed physics mechanism of ridge formation.

\section{Outlook}
STAR is actively pursuing new analysis procedures that attempt to address specific questions on medium
properties, mainly by using multi-hadron
correlations.
First results on di-jet triggered di-hadron correlations were reported~\cite{olga}. The results revealed features 
very different from those reported for single high $p_{T}$ triggered correlations. There is no 
observation of away-side yield suppression, away-side shape modification or near-side ridge formation 
for the di-jet triggered events. A detailed study of these events with clever changes in trigger and 
associated particle $p_{T}$ ranges can provide a controlled tool to further understand jet-medium 
interactions in heavy-ion collisions.	The inferred elliptic flow from events where two high $p_{T}$ 
particles lie in the ridge show that it could be  higher than for events where two high $p_{T}$ particles lie 
in the jet cone~\cite{paul}. 
This study holds the potential to provide further constraints on understanding the physics mechanism of ridge 
formation.	
STAR also started its first attempts towards full jet reconstruction in heavy ion collisions by reporting an 
analysis based on multi-hadron cluster triggered correlations~\cite{brooke}.

\end{document}